\documentclass[12pt,preprint]{aastex}

\shorttitle{A POSSIBLE BROWN DWARF COMPANION\\ TO THE WHITE
 	DWARF GD1400}
\shortauthors{J. Farihi}

\begin{document}

\title{A POSSIBLE BROWN DWARF COMPANION\\ TO THE WHITE
 	DWARF GD1400}

\author{J. Farihi}
\affil{Department of Physics \& Astronomy, University of California, 
	Los Angeles, CA 90095}
\email{jfarihi@astro.ucla.edu}

\and

\author{M. Christopher}
\affil{Department of Astronomy, California Institute of Technology,
 	Pasadena, CA 91125}
\email{mc@astro.caltech.edu}

\begin{abstract}

An unresolved, likely L dwarf companion to the DA white dwarf
GD1400 is reported.  This would be only the second such system
known, discovered 17 years after the prototype L dwarf, GD165B,
was determined to be a companion to a white dwarf.  Photometric
observations and model predicted stellar parameters of the well
studied white dwarf primary indicate that GD1400B has $J-K\ga2.0$
and $M_K=12.13$ mag.  If correct, this would place GD1400B at
spectral type L6, and it would be the lowest luminosity unevolved
companion known to a white dwarf, and thus a definite brown dwarf.  
However, a low resolution Keck NIRSPEC $2.1-2.4\mu$m spectrum may
not be consistent with known L dwarfs.  Uncertainties in
classification remain until the binary is resolved or a
trigonometric parallax is measured.

\end{abstract}


\section{INTRODUCTION}

In 1988, Becklin \& Zuckerman announced the discovery of the
coolest dwarf star yet seen.  Imaged with the first generation of
infrared cameras, this companion to the white dwarf GD165 went
without proper classification for several years.  The L dwarf
prototype, GD165B came to represent the first new class of
main sequence stars in 100 years, a class believed to retain
both stellar and substellar objects over the age of the Galaxy
\citep{bur97,kir99,kir00,burg03}.  Since that time, over two
hundred and fifty objects of this spectral type have been
discovered.  Rather ubiquitous in the field and in young clusters,
L dwarfs have remained relatively infrequent companions to stars
with $M\ga0.2M_{\odot}$ \citep{zuc87,zuc92,sch00,opp01,hin02,far03,
mcc04}.  Understanding the overall statistics and individual
properties of the lowest mass companion systems is important
for both star and planet formation.

This paper describes the properties of a very cool companion
to the white dwarf GD1400 (WD0145-221, ${\rm 01^h\ 47^m\ 21.8^s,
-21\arcdeg\ 56'\ 51.4''}$ Eq. J2000).  Spectroscopic and photometric
evidence is presented that indicates the presence of a low mass, L
type dwarf within $0.3''$ of the primary.

\section{OBSERVATIONS \& DATA}

\subsection{\it $V$ Band Photometry}

Optical $V$ band data were taken 3 January 2004 with the CCD
Camera on the Nickel 1 meter telescope at Lick Observatory.
GD1400 and a nearby Tycho 2 catalog star (TYC4688-111-1) were
both observed for a total of 3 minutes, each in 3 separate 1
minute exposures, yielding ${\rm SNR}\sim240$ on GD1400 and
${\rm SNR}\sim7000$ on the $V=11.24$ mag calibrator
\citep{hog00,bes00}.  Conditions were clear but with poor
seeing ($\theta\sim4.5''$) that appeared to remain relatively
stable between target and calibrator observations.  The individual
frames were bias subtracted, flat fielded, registered and then
averaged to create a single reduced image upon which to perform
photometric measurements.

Photometry was executed with $10''$ and $20''$ diameter
apertures on both GD1400 and calibrator, including telluric
extinction corrections, producing consistent results to within
$0.01$ mag.  The error in the measured flux of each object was
$<0.01$ mag.  The published uncertainty in the magnitude of
TYC4688-111-1 is 0.12 mag \citep{hog00}.  The result for
GD1400 is $V=14.85\pm0.12$ mag. 

\subsection{\it $K$ Band Spectroscopy}

GD1400 was observed on 16 January 2004, using the Near
Infrared Spectrograph (NIRSPEC, McLean et al.\ 1998) at Keck
Observatory.  NIRSPEC was used in low resolution mode with a
$0.76''$ slit, yielding a spectral resolution of $\lambda/\Delta
\lambda\sim1500$.  The seeing was poor during the observations
($\theta\sim1.2''$ at $K$) and the telescope suffered from
significant windshake at the location of GD1400, so no
companion was resolved, nor was any elongation detected
except due to the effects of windshake.

Two 10 minute spectra (dithered by $15''$ for sky subtraction)
were obtained in $K$ band using the N7 filter, which yielded a
wavelength coverage of $2.04-2.45\mu$m.  The observations had a
fixed position angle on the sky of $38\arcdeg$.  HD63586, an A0V
star, was observed at similar airmass for atmospheric calibration.

The spectrum of GD1400 was reduced using the REDSPEC data
reduction package.  The images were cleaned of bad pixels,
flat fielded, pair subtracted, and spatially rectified.  Two
spectra of GD1400 were extracted, averaged, divided by the
calibrator spectrum (scaled by a blackbody), wavelength
calibrated and normalized.  The average of the two spectra
is shown in Figure \ref{fig1}.

\section{RESULTS}

An unresolved companion was suspected based on the optical
and near infrared data available on GD1400.  In the following
sections, evidence is presented that reveals a low mass companion
causing excess emission at $2.2\mu$m.

\subsection{\it J \& K Magnitudes and Color}

There exist two independent spectroscopic analyses of the hydrogen
rich (DA) white dwarf GD1400.  Both \citet{koe01} and \citet{fon03}
give results based on high resolution spectroscopy that agree quite
well; $T_{\rm eff}=11,605$ K, log $g=8.05$ and $T_{\rm eff}=11,550$
K, log $g=8.14$ respectively.  The average and standard deviation of
these two results ($11,580\pm40$ K, log $g=8.10\pm0.06$) were used
to predict model colors and absolute magnitudes, with errors, for
GD1400 \citep{ber95}.

Before considering distances and absolute magnitudes for
GD1400B, the flux ratio of primary and secondary should be
considered.  Based on the measured $V$ magnitude and the $V-K=
-0.24$ model predicted color for GD1400, the predicted $K$ magnitude
is $K=15.09\pm0.12$ mag (model predicted color has less than 0.01 mag
uncertainty and is essentially independent of surface gravity -- the
error is all measurement error in $V$).  The 2MASS point source
catalog gives a SNR=19.2 measurement of $K_s=14.34\pm0.06$ mag for
the composite pair.  Ignoring corrections between $2.2\mu$m filter
sets, a deconvolved magnitude of $K=15.10\pm0.20$ mag is calculated
for the companion.  If the white dwarf models are accurate, the
companion and primary appear equally luminous at this wavelength.

There is little or no excess emission measured for GD1400 at
$1.2\mu$m, which is consistent with a very red color of $J-K\ga2.0$
for its companion (Table \ref{tbl-1}); consistent with a spectral
type of L5.5 or later \citep{kir00,dah02}.  Within the uncertainties
of the 2MASS measured $J$ magnitude of the composite and the white
dwarf model extrapolation, it is possible that the companion $J$
band flux is undetected.

The radius of GD1400, which has been determined spectroscopically
by two independent measurements, can be used to estimate an absolute
magnitude for GD1400B.  At an assumed 39 pc photometric distance for
GD1400 (based on the model predicted absolute magnitude of $M_V=11.88
\pm0.10$ mag for the white dwarf) an absolute $K$ magnitude of $M_K=12.13
\pm0.22$ mag is determined for the L dwarf companion.  These data
are most compatible with spectral type L6 and $T_{\rm eff}\sim1650$
K for GD1400B \citep{rei99,kir00,dah02}.

\subsection{\it A Comparison with the GD165 Binary System}

Coincidentally, the GD165 system (DA4.2+dL4), which has a white
dwarf primary with similar temperature and surface gravity to GD1400,
will serve as a useful guide and comparator for the following analysis.
There are two independent spectroscopic parameter determinations for
GD165.  Both \citet{koe01} and \citet{ber95b} give similar results;
$T_{\rm eff}=11,970$ K, log $g=7.91$ and $T_{\rm eff}=11,980$ K,
log $g=8.06$, respectively.  Also, GD165 has a trigonometric parallax
which places it at 31.5 pc.  The difference between the model predicted
versus measured absolute magnitudes for GD165A are 0.06 mag at $V$,
0.03 mag at $J$ and 0.04 mag at $K$ \citep{ber95,ber95b,bec88}.
Hence the models appear quite accurate in this case.

In contrast to the equally luminous components of the GD1400
system, GD165B is 0.41 mag brighter than its white dwarf primary
at $K$.  Provided that brown dwarfs and the lowest mass stars all
have radii near $1R_{\rm J}$ after 1 Gyr \citep{bur01}, GD1400B
appears cooler than the L4 dwarf GD165B.  The total $J$ band flux
of the GD1400 system is consistent (within the errors) with a single
star, yet the combined $J$ band flux of both components of the GD165
system is 0.29 mag brighter than the flux from the white dwarf alone 
\citep{ber95}.  Again, if the white dwarf and brown dwarf models used
here are accurate, GD1400B appears to be cooler than GD165B.

Although this photometric evidence points to an object cooler
than GD165B, there is still room for some error in the models
and perhaps even in the spectroscopy of the primary white dwarf.
The GD1400 system is unresolved and there currently exists no
trigonometric parallax.

\subsection{\it $K$ Band Spectrum}

\begin{figure}
\plotone{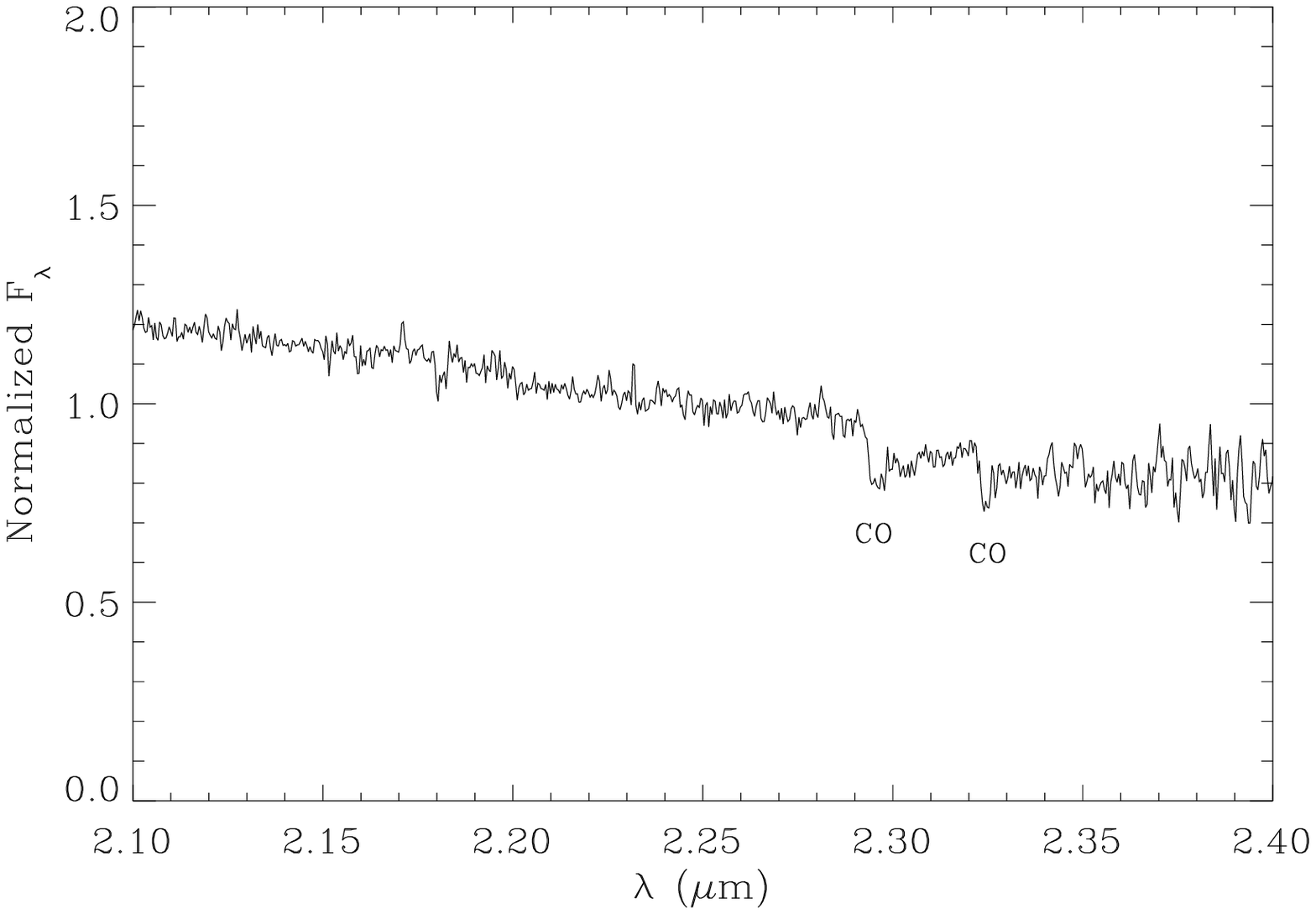}
\caption{Combined spectrum of GD1400 and L dwarf companion,
normalized to 1.0 at $2.235\mu$m
\label{fig1}}
\end{figure}

\begin{figure}
\plotone{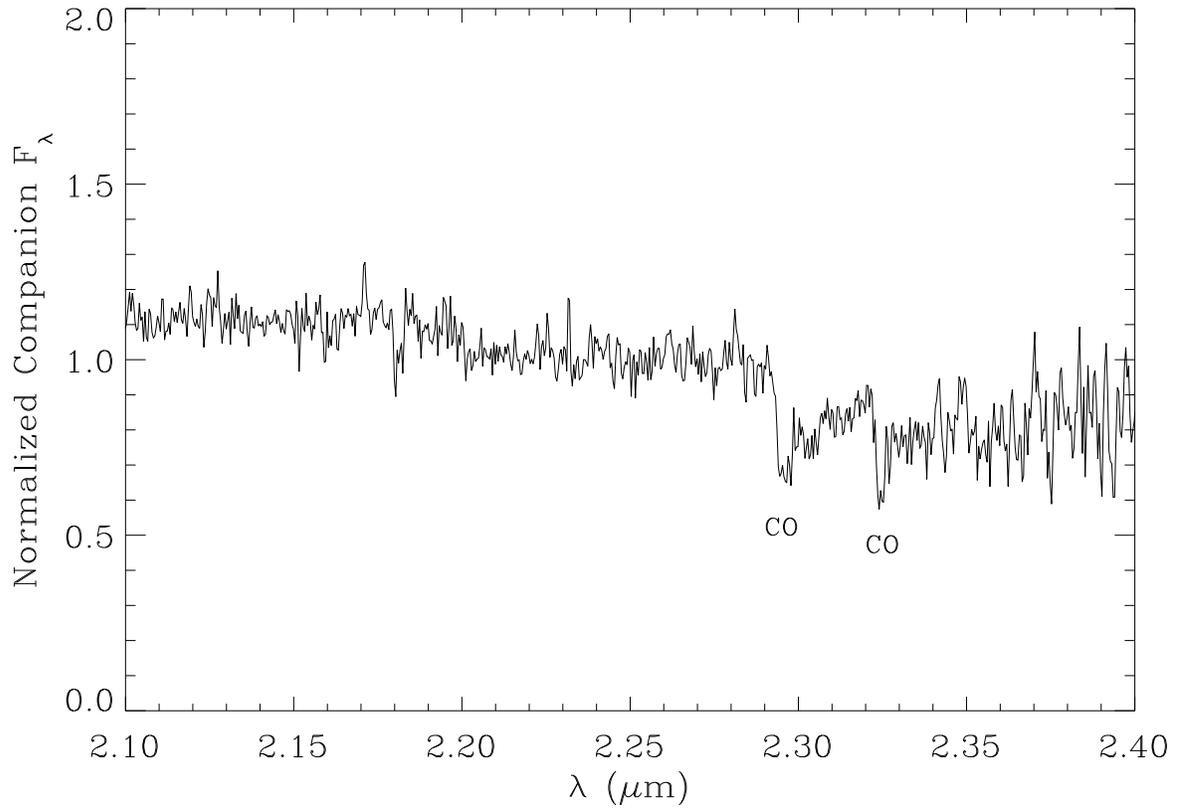}
\caption{L dwarf companion spectrum after subtraction of
a 11,500 K blackbody of equal flux at $2.2\mu$m.  The CO bandheads
at $2.29\mu$m and $2.32\mu$m are clearly seen.  Also note the
absence of Na at $2.21\mu$m.  The standard deviation between
$2.07-2.28\mu$m is 0.06 flux units per resolution element.
\label{fig2}}
\end{figure}

Near infrared spectroscopy was performed to examine the
possibility of a red background object or dust as the cause of the
excess emission seen at $2.2\mu$m.  The $2.1-2.4\mu$m spectrum of GD1400
is shown in Figure \ref{fig1}.  The two features seen at $2.29\mu$m and
$2.32\mu$m are CO bandheads, which are indicative of a low mass companion
and readily visible in this composite spectrum.  An 11,500 K blackbody
spectrum, scaled to have exactly half the flux of the composite at
$2.20\mu$m, was subtracted from the spectrum in Figure \ref{fig1},
thus effectively removing the expected contribution from the
featureless white dwarf at these wavelengths.  The resulting
spectrum is shown in Figure \ref{fig2}.

\begin{figure}
\plotone{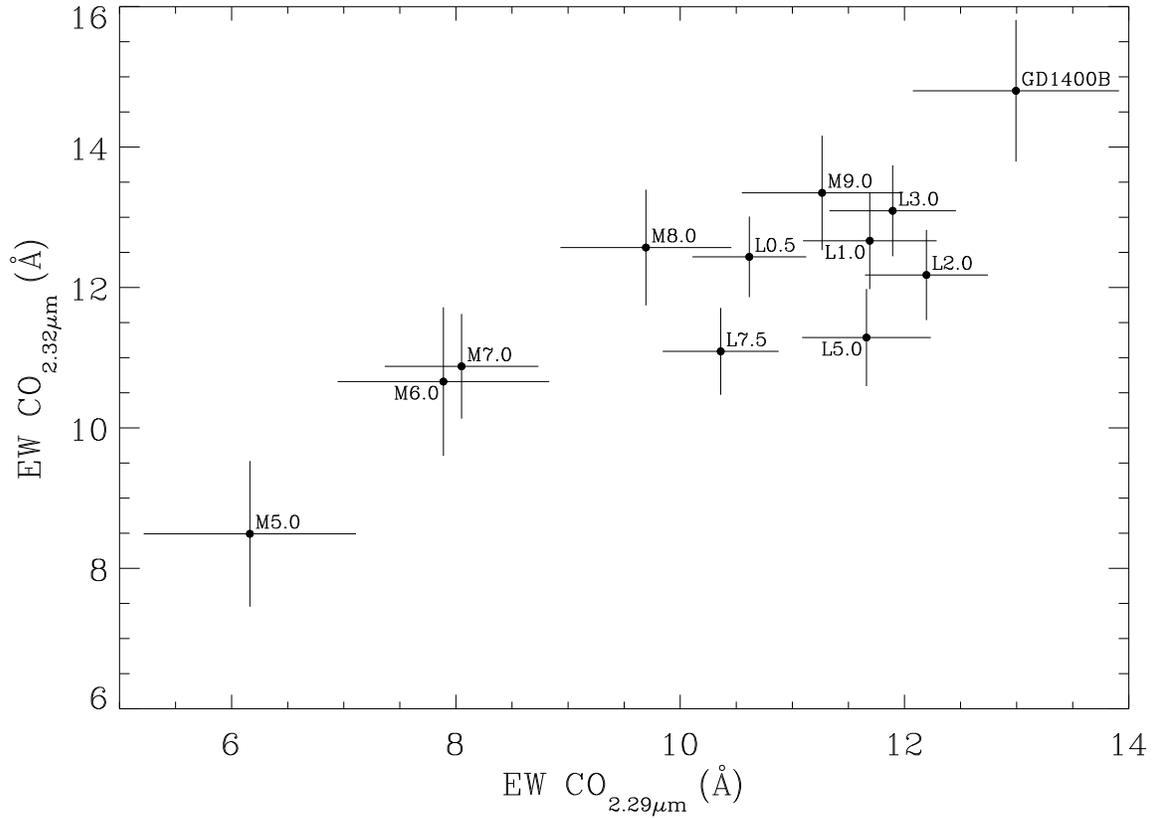}
\caption{Equivalent widths measured for individual M and L
dwarfs from \citep{cus00}.  GD1400B appears somewhat anomalous.
\label{fig3}}
\end{figure}

In the subtracted spectrum, both CO bandheads have ${\rm SNR}>25$.
Figure  \ref{fig3} shows the equivalent widths of these features for
several M and L dwarfs from \citet{cus00} together with GD1400B --
all measured in a uniform way for this work.  The equivalent widths
for GD1400B appear stronger than those of known single L and M dwarfs.
Oversubtraction of the white dwarf continuum from the composite spectrum
could, in principle, cause the measured CO bandhead equivalent widths to
be larger than their intrinsic values.  Another possibility is close
binary interactions, such as heating of the companion by the white
dwarf.  Further observations are necessary to reconcile this
apparent anomaly.

There is an absence of Na at $2.21\mu$m, which can be seen weakly
as late as spectral type L2, but not later.  Equivalent widths vary
between $0.2-3.0$ \AA{} for early L dwarfs \citep{mcl03,cus00}.  A
measurement of this region in the spectrum of GD1400B is consistent
with an equivalent width of $<1.0$ \AA{}.  The noise in the region from
$2.19-2.23\mu$m is 0.040 flux units per resolution element and the
Na line typically has a full width of $\sim50$ \AA{} \citep{mcl03,
cus00} in late M dwarfs, or roughly 5 resolution elements of 2.6 pixels
each.  Hence, an equivalent width as small as 1.0 \AA{} could have
been detected at the $1\sigma$ level.  The lack of Na is inconsistent
with an M dwarf.

The conclusion based on spectroscopy alone is that an early
L type cannot be ruled out with confidence.  Although the
equivalent widths of the first and second CO bandheads are larger
than those seen in single L and late M dwarfs, this could be due
to close binary interactions or model uncertainties which led to
oversubtraction of the white dwarf flux.  The fact that Na is not
manifest in the spectrum may be at odds with the CO equivalent
widths, but is consistent with the photometry.

\section{DISCUSSION}

\subsection{\it The Age \& Mass of GD1400B}

GD1400, a $0.67M_{\odot}$ white dwarf, has been cooling for 0.51
Gyr according to models (P. Bergeron 2002, private communication). 
From the initial to final mass relation for white dwarfs \citep
{wei87,wei90,wei00,bra95}, a white dwarf with this mass has evolved
from a main sequence progenitor with a mass in the range $2.5M_
{\odot}-3.0M_{\odot}$.  The main sequence lifetime then
falls within the range $0.5-1.0$ Gyr \citep{mae89}, yielding
a total age between $1.0-1.5$ Gyr.  The very moderate proper
motion of GD1400, $\mu\approx0.05'' \ {\rm yr^{-1}}$ \citep{zac04},
also supports the conclusion that this is a relatively young system.
Assuming $v_{rad}=0$, its Galactic space velocity relative to
the LSR is $(U,V,W)=(-4,+5,+9) \ {\rm km \ s^{-1} }$ ($U$ is taken
to be positive toward the Galactic anticenter, $V$ positive in the
direction of Galactic rotation, and $W$ positive toward the North
Galactic Pole), with a total space motion of only $9 \ {\rm km \
s^{-1} }$ with respect to the Sun -- typical of young disk objects
\citep{mih81,leg92}.  At 1.25 Gyr, a $T_{\rm eff}\sim1650$ K brown
dwarf is predicted to have $M\approx60M_{\rm J}$ \citep{cha00}.

\subsection{\it Origin \& Evolution of the GD1400 Binary System}

The orbital separation of GD1400B is uncertain.  The system
remained unresolved in subsequent NIRSPEC $K$ band images taken
two weeks after the spectrum was obtained.  The image with the
best spatial resolution reveals a full width of $\theta=0.82''$
and a sufficiently symmetric PSF to rule out an equally luminous
binary with separation $\ga0.3''$.  Although there is the
possibility of a chance alignment, the most likely explanation
for the unresolved nature of the binary is a relatively small,
$a<12$ AU, semimajor axis.

During the AGB phase of stellar evolution, orbiting bodies outside
of $\sim5$ AU should eschew the expanding envelope and eventually
have their semimajor axes expanded by a factor (typically 3 to 5)
related to the mass lost \citep{jea24,zuc87}.  Inside of $\sim5$ AU,
a companion should inspiral due to tidal and frictional interaction
with the slow AGB wind and expanding photosphere \citep{deb02}.
Hence, it is likely that the semimajor axis of GD1400B's orbit
is less than a few AU.  A sensitive search for radial velocity
variations might be able to confirm or rule out a close binary
system.

\section{CONCLUSION}

Based on photometric data presented here, the first unambiguous
substellar companion to a white dwarf may have been discovered within
$0.3''$ ($\sim12$ AU) of GD1400 (DA4.4).  However, the near infrared
spectroscopy may indicate a CO temperature that is higher than the
photospheric temperature indicated by the photometry.  This
could be the result of close binary interactions causing the CO bandheads
to appear more strongly than in isolated L dwarfs or may be due to
errors in the models used to remove the flux of the primary.

From a direct comparison with the GD165 binary system (relying on the
same white dwarf and brown dwarf models), it appears that GD1400B is
cooler than spectral type L4.  The deconvolved magnitudes and colors
of GD1400B are most consistent with spectral type L6, indicative of
a $60M_{\rm J}$ brown dwarf at an estimated age of 1.25 Gyr.  Its
true nature and origins remain uncertain until more is known about
the binary system.  A trigonometic parallax determination,
radial velocity measurements, or very high resolution imaging
should be able to further constrain the properties of this
system.

\acknowledgments

Both authors owe a debt of gratitude to M. Cushing for kindly
providing us with his spectra for analysis here and to N. Scoville
for donating some of his Keck NIRSPEC time to observe GD1400.  J.
Farihi wishes to express sincere thanks to B. Zuckerman for his
assistance in acquiring the optical data, to E. Gates of Lick
Observatory for taking the optical data, to L. Prato \& M. McGovern
for securing follow up near infrared images and many helpful
discussions, and to E. E. Becklin \& B. Zuckerman for careful
readings of the manuscript and constructive comments.  Some of
the data presented herein were obtained at Keck Observatory,
which is operated as a scientific partnership among the California
Institute of Technology (CIT), the University of California
and the National Aeronautics and Space Administration (NASA).
Some data used in this paper are part of the Two Micron All
Sky Survey, a joint project of the University of Massachusetts
and the Infrared Processing and Analysis Center (IPAC)/CIT,
funded by NASA and the National Science Foundation (NSF).
2MASS data were retrieved from the NASA/IPAC Infrared Science
Archive, which is operated by the Jet Propulsion Laboratory,
CIT, under contract with NASA.  J. Farihi has been supported
in part by grants from NASA to UCLA and M. Christopher by
NSF grant AST02-28955.
 
Facilities: \facility{Nickel}, \facility{Keck}.


\clearpage

\clearpage

\clearpage

\begin{deluxetable}{lcccc}
\tablecaption{Magnitudes for GD1400 and Companion \label{tbl-1}}
\tablewidth{0pt}
\tablehead{
\colhead{Filter}			&
\colhead{$\lambda_{0}$ ($\mu$m)}	&
\colhead{Composite}			&
\colhead{White Dwarf (mag)}		&
\colhead{L Dwarf (mag)}}

\startdata

$V$	& 0.55	&$14.85\pm0.12$	&$14.85\pm0.12$	& \nodata\\
$J$	& 1.22	&$14.92\pm0.03$	&$15.02\pm0.12$	
				&$17.5\pm0.9$\tablenotemark{\dag} \\
$H$	& 1.65	&$14.45\pm0.05$	&$15.02\pm0.12$	&$15.42\pm0.26$ \\
$K$	& 2.16	&$14.34\pm0.06$	&$15.09\pm0.12$	&$15.10\pm0.20$ \\

\enddata

\tablecomments{Measured magnitudes for the composite star are
from this work and 2MASS ($2.2\mu$m data point is a $K_s$
magnitude).  Predicted near infrared magnitudes for
the white dwarf are from \citet{ber95}.}

\tablenotetext{\dag}{Deconvolved L dwarf $J$ magnitude is extremely
sensitive to small changes in the measured and predicted magnitudes
(\S3.1).}

\end{deluxetable}

\end{document}